\documentstyle[aps,epsf,rotate,multicol]{revtex}

\begin{document}

\draft

\title{Scale-invariant Truncated L{\'e}vy Process}  

\author{Boris Podobnik$^{1,2}$, Plamen~Ch.~Ivanov$^{2,3}$, Youngki Lee$^{2}$, 
and H.~Eugene~Stanley$^2$}

\address{
$^1$Department of Physics, Faculty of Science, University of Zagreb, Zagreb, 
Croatia\\
$^2$Center for Polymer Studies and Department of Physics,
  Boston University, Boston, MA 02215\\
$^3$Cardiovascular Division, Harvard Medical School, Beth Israel
  Hospital, Boston, MA 02215 }

\maketitle

\maketitle

\begin{abstract}

We develop a scale-invariant truncated L{\'e}vy (STL) process to
describe physical systems characterized by correlated stochastic
variables. The STL process exhibits L{\'e}vy stability for the
probability density, and hence shows scaling properties (as observed
in empirical data); it has the advantage that all moments are finite
(and so accounts for the empirical scaling of the moments).  To test
the potential utility of the STL process, we analyze financial data.

\end{abstract}

\begin{multicols}{2}

In recent years, the L{\'e}vy process \cite{L37} has been proposed to
describe the statistical properties of a variety of complex phenomena
\cite{Shl93,Sol,Ott,Bar,Hay94,Moo,Zum94,Orr90}. The L{\'e}vy process 
is characterized by ``fat tails'' (power law), 
and display scaling behavior similar to that observed in a wide
range of empirical data.  However, the application of the L{\'e}vy
process to empirical data is limited because it is characterized by
infinite second and higher moments, while empirical data have finite
moments.

Truncated L{\'e}vy (TL) processes are defined to have a L{\'e}vy
probability density function (PDF)  in the central regime, truncated
by a function decaying faster than a L{\'e}vy distribution in
the tails \cite{Mantegna94}.  The TL process is introduced to account
for the finite moments observed for empirical data \cite{MS95}. However,
the TL process (with either abrupt \cite{Mantegna94} or smooth
\cite{Koponen95} truncation) has limitations when applied to empirical
data. (i) The TL process is introduced for independent and identically
distributed (i.i.d.) stochastic variables, while variables describing
many physical systems are not i.i.d. --- e.g. there are correlations. 
(ii) The PDF of the
TL process tends to the Gaussian distribution (according to the
central limit theorem), and hence does not exhibit scale invariance;
PDFs for a variety of complex systems, however, are often
characterized by regions of scale-invariant behavior. (iii) The
time scale above which the L{\'e}vy profile becomes Gaussian depends
on the size of the truncation cutoff (or the standard deviation)
\cite{Mantegna94,Koponen95}; to mimic the L{\'e}vy type scale
invariant behavior observed for data, the TL process must be defined
with a standard deviation larger than the one observed for the
data [see caption to Fig.1].

Here we introduce a type of stochastic process which we call the
scale-invariant truncated L{\'e}vy (STL) process.  
Stochastic variables $z$ in the STL are generated 
by the symmetrical
probability function $f(z) = A e^{-\lambda |z|^{\beta}}
|z|^{-1-\alpha}$, where $0<\alpha<2$ \cite{footnote0}. 
The exponential prefactor
\cite{Koponen95} ensures a smooth truncation of the L{\'e}vy
distribution, where the parameter $\beta$ can take any positive value
\cite{footnote1}, $\lambda^{-1}$ is related to the size of the
truncation cutoff, and $A$ is a measure of the ``spread'' in the
central region.

From the probability function $f(z)$, we calculate the characteristic
function $\phi(k) \equiv \exp[-\int^{\infty}_{-\infty}dz(1-e^{-ikz})f(z)]$ 
\cite{GK54}.  The PDF ${\cal P}_{\Delta t}(z)$ is
the Fourier transform of $\phi(k)$ \cite {GK54}:
\begin{equation}
{\cal P}(z) \equiv
\frac {1}{2\pi} \int \phi(k)
e^{ikz} dk,
\label{probability}
\end{equation}
Since $f(z) \approx A|z|^{-1-\alpha}$ for small values of $z$, ${\cal
P}(z)$ has a L\'{e}vy profile in the central part.  To maintain scale
invariance for ${\cal P}(z)$ in the entire range including the tails
\cite{footnote2}, we introduce the scaling transformations
\begin{equation}
A_{\Delta t} \equiv ({\Delta t})^{\epsilon} A_1,~~~~~
\lambda_{\Delta t} \equiv ({\Delta t})^{-\epsilon \beta/\alpha} \lambda_1,
\label {scale-invar.tr}
\end{equation}
where $\Delta t$ is the time scale and $\epsilon $ can take any
positive value.  Under these transformations, the PDF ${\cal
P}(z)={\cal P}_{\Delta t}(z)$ scales as the L{\'e}vy stable
distribution:
\begin{equation}
z \equiv ({\Delta t})^{\epsilon/\alpha} z_1,~~~~~~
{\cal P}_{\Delta t}(z) \equiv
\frac{{\cal P}_1(z_1)}{({\Delta t})^{\epsilon/\alpha}}.
\label {prob.scale-invar}
\end{equation}
With the transformations of Eqs.~(\ref{scale-invar.tr}) and
(\ref{prob.scale-invar}), we obtain a process with controlled
dynamical properties --- ${\cal P}_{\Delta t}(z)$ for any value of 
$\Delta t$ can be calculated from the PDF at any chosen $\Delta t$
(e.g. $\Delta t=1$) \cite{footnote3}.

Although the PDF ${\cal P}_{\Delta t}(z)$ exhibits scale invariant
properties identical to the L\'{e}vy stable distribution, the process
defined by Eqs.~(\ref{probability})~and~(\ref{scale-invar.tr}) is
different.  While the L\'{e}vy process is defined for i.i.d. variables
the STL process is characterized by correlated stochastic variables
--- the STL is a non-i.i.d. type process. To demonstrate this, we
consider the scaling of the second moment $\sigma^2$, determined as
the second derivative of $\phi(k)$ at small values of $k$ \cite{GK54}:
\begin{equation}
\sigma^2_{\Delta t}=~\frac {2 A  ~\Gamma ((2-\alpha)/\beta) 
~\lambda ^{(\alpha -2)/\beta}}{\beta}=
({\Delta t})^{2 \epsilon/\alpha} \sigma_1^2,
\label {sigma-scale}
\end{equation}
where $\sigma_1$ is the initial standard deviation for $\Delta t=1$.
The second equality on the right hand side follows from the
transformations of Eq.~(\ref{scale-invar.tr}).  For an appropriate
choice of $\epsilon/\alpha~ (\neq0.5$), the scaling relation
(\ref{sigma-scale}) indicates the presence of correlations that can be
positive (or negative).  In addition, the STL process exhibits scaling
not only for the second moment but for all higher moments:
\begin{equation}
< | z | ^n > \equiv \int dz~| z | ^n~{\cal P}_{\Delta t}(z)
= {\Delta t}^ {\epsilon n /\alpha}  < | z_1 | ^n > .
\label{moments}
\end{equation}

Hence, the STL is a process for which the PDF ${\cal P}_{\Delta
t}(z)$, the second moment $\sigma^2$, and all higher moments 
$< | z |^n >$ scale with the same scaling exponent $\epsilon/\alpha$.

Often with empirical data, we observe several different scaling
regimes.  To account for a crossover at given time scale $(\Delta
t)_{\times}$, we introduce different scale invariant transformations
from the type of Eq.~(\ref{scale-invar.tr}) for two different regimes
of time scales:
\begin{mathletters}
\begin{equation}
\lambda_{\Delta t} =
\left \{ \begin{array}{c}
(\Delta t)^{-{\epsilon}_1 \beta/\alpha} \lambda_1~~~~~~~ 1 \le \Delta t
\le (\Delta t)_{\times}
\nonumber\\
(\Delta t)^{-{\epsilon}_2 \beta/\alpha} \lambda_{\times}~~~~~~~~~~~ \Delta t >
(\Delta t)_{\times} \end{array}
\right \},
\label{scale2}
\end{equation}
\begin{equation}
A_{\Delta t} =
\left \{ \begin{array}{c}
(\Delta t)^{\epsilon_1} A_1~~~~~~~~~~~~~1 \le \Delta t \le (\Delta t)_{\times}
\nonumber\\
(\Delta t)^{\epsilon_2} A_{\times}~~~~~~~~~~~~~~~~~~\Delta t > (\Delta t)_{\times}   \end{array}
\right \}.
\label{scale3}
\end{equation}
\end{mathletters}
Here $\alpha$, $A_1$ and ${\lambda}_1$ are free parameters, chosen to
fit ${\cal P}_{\Delta t}(z)$ at the initial time scale $\Delta t = 1$.
Continuity of the PDF and the moments between the two
scaling regimes is ensured by continuity in the values of $A$ and $\lambda$: 
from Eqs.~(\ref{scale2}) and (\ref{scale3}) we find 
$A_{\times} \equiv (\Delta t)_{\times}^{{\epsilon}_1-{\epsilon}_2} A_1$ 
and $\lambda_{\times} \equiv (\Delta
t)_{\times}^{\beta ({\epsilon}_1-{\epsilon}_2)/\alpha} \lambda_1 $.

To exemplify the features of the STL process for a broad range of time
scales, we need sufficiently large data sets. Such a large data set is
the $S\&P500$ stock index over the 12 year period Jan '84-Dec '95.
The price fluctuations $z$ of this index are the stochastic variable
analyzed.  In particular, we focus on the scaling behavior of several
statistical characteristics: (1) the second and higher moments, (2)
the probability of return to the origin ${\cal P}_{\Delta t}(0)$, and
(3) the PDF ${\cal P}_{\Delta t}(z)$.  For
simplicity we set $\beta=1$.

We make three empirical observations.
(i) Experimental results for the standard deviation as a function of 
$\Delta t$ show two different scaling regimes with a crossover at 
$(\Delta t)_{\times} \approx 30$~min \cite{MS95} [Fig.1]. 
The regime at small time
scales is characterized by slope $0.7$, indicating the presence of positive 
correlations in the price fluctuations $z$ (``superdiffusive'' regime). The
second regime has slope $0.5$, indicating absence of correlations 
(``normal diffusion'' regime). Therefore the 
fluctuations in the $S\&P500$ index cannot be described by an
i.i.d. stochastic process, such as the L{\'e}vy or the TL process.
(ii) The probability of return 
to the origin ${\cal P}_{\Delta t}(0)$, however, exhibits a L{\'e}vy type of
scaling for more than three decades [Fig.2]. Such scaling 
for ${\cal P}_{\Delta t}(0)$ therefore indicates L{\'e}vy
scale invariance of the central part of the probability density.
(iii) The scaling exponent of ${\cal P}_{\Delta t}(0)$
is identical to the exponent of the standard deviation in the first scaling
regime. However, the crossover in the scaling of the standard deviation
is not followed by a change in the slope of ${\cal P}_{\Delta t}(0)$.

To account for the first empirical observations, we introduce a
stochastic process with two different regimes: 
(a) a STL regime with $A_{\Delta t} \equiv ({\Delta t})^{\epsilon} A_1$ and
$\lambda_{\Delta t} \equiv ({\Delta t})^{- \epsilon/\alpha} \lambda_1$, to
account for the superdiffusive behavior $\sigma \propto (\Delta
t)^{\epsilon/\alpha}$ (Eq.~\ref{sigma-scale}) at short time scales
$\Delta t<(\Delta t)_{\times}$ [Fig.1]; and 
(b) a regime with breakdown of scaling defined by $\lambda_{\Delta t} \equiv
\lambda_{\times}=const$ and 
$A_{\Delta t} \equiv ({\Delta t}) A_{\times}$ for $\Delta
t>(\Delta t)_{\times}$ to account for the normal diffusive behavior
$\sigma \propto (\Delta t)^{1/2}$ (Eq.~(\ref{sigma-scale}) and Fig.1).  
This breakdown allows for a transition from a non-i.i.d. STL process to
an i.i.d. TL process.

The STL process in the regime $\Delta t<(\Delta t)_{\times}$ 
accounts for the second empirical observation, the
identical scaling exponent
($\epsilon/\alpha$) experimentally observed for both the standard
deviation $\sigma$ (Eq.~\ref{sigma-scale}) 
and the probability of return to the origin ${\cal
P}_{\Delta t}(0)$ (Eq.~\ref{prob.scale-invar} and Fig.2).  
From fitting the initial probability
distribution ${\cal P}_1(z)$, we obtain $\alpha=1.43$. Since
empirically the standard deviation scales with exponent
$\epsilon/\alpha=0.7$, we find that $\epsilon=1$ for this process.

Third, we find that the theoretical prediction for the STL process
with a scaling breakdown is in good agreement with the empirical
result for ${\cal P}_{\Delta t}(0)$ for more than three decades
[Fig.2].  We note that the transition at $(\Delta t)_{\times}\approx
30$ from STL (non-i.i.d.) process to a TL (i.i.d.) process in the
scaling of $\sigma$ [Fig.1], does not imply a sharp transition in the
scaling of ${\cal P}_{\Delta t}(0)$ from a L\'{e}vy to Gaussian
behavior [Fig.2].  The reason is that the STL scaling regime
(Eq.~(\ref{scale-invar.tr})), ${\cal P}_{\Delta t}(0)$ exhibits
L\'{e}vy scaling behavior (Eq.~(\ref{prob.scale-invar})) up to
$(\Delta t)_{\times}\approx 30$.  In this scaling regime, $\sigma$
increases superdiffusively with exponent 0.7, that is much faster than
0.5 for an i.i.d. process. At the crossover scale $(\Delta
t)_{\times}$, the standard deviation reaches the value
$\sigma_{\times}=(\Delta t)_{\times}^{0.7} \sigma_1$.  The value of
$\sigma_{\times}=(\Delta t)_{\times}^{0.5} \sigma_{TL}$ can be also
related to an i.i.d. TL process with initial standard deviation
$\sigma_{TL}>\sigma_1$ [Fig.1].  
According to the central limit
theorem, an i.i.d. TL process asymptotically converges to a Gaussian
process. Thus while in the short time regime (small $\Delta t$) the
price fluctuation $z$ over time $\Delta t$ is a sum of correlated
stochastic variables, in the asymptotic regime (large $\Delta t$), $z$
can be treated as a sum of newly-defined independent stochastic
variables with standard deviation $\sigma_{TL}$.  Since such a
Gaussian process is defined with large initial standard deviation
$\sigma_{TL}$, the transition from the L{\'e}vy to the Gaussian
behavior is delayed [Fig.2].  
The time scale $(\Delta t)_s$ of this
transition can be calculated by equating the return probability ${\cal
P}_{\Delta t}(0)$ for the L{\'e}vy and Gaussian distributions
\cite{footnote4}. We find that $(\Delta t)_s={\cal B}(\Delta
t)_{\times}$, where ${\cal B}\approx 70$ [Fig.2]. Such a relation is
interesting, since it explicitly connects the crossover from the
L{\'e}vy to Gaussian with the crossover from non-i.i.d. to
i.i.d. process.

Finally, we compare the empirical distributions of the price
increments $z$ of the $S\&P500$ index for different time scales $\Delta t$
with the shape of the distributions obtained analytically [Fig.3].
Good agreement between data and the theoretical distributions is
observed both for the central part and for the tails.  At small time
scales, the scale-invariant behavior of ${\cal P}_{\Delta t}(z)$ is
maintained in the entire range (L{\'e}vy for the central profile, and
exponential in the tails) due to the scaling transformations 
of the STL process (Eq.~\ref{scale-invar.tr}). 
The crossover to an
i.i.d. TL process at large time scales ensures a smooth transition to a
Gaussian-like profile. We find that the proposed mechanism of a STL
process, with breakdown, provides a reliable control of the dynamical
properties of the PDF.

In our analysis, we have considered the price fluctuations of the 
$S\&P500$ index as the stochastic variable $z$. The choice of stochastic
variable depends on the type of the stochastic process: e.g., for an
{\it additive} process one considers increments, while for 
{\it multiplicative} 
processes the appropriate choice is relative increments. In finance, it
is traditionally assumed that economic indicators arise from a
multiplicative process, and correspondingly the preferred quantity to 
analyze is the rate of return or the difference in the natural logarithm
of price. The additive and multiplicative processes are related 
for high frequency data (small $\Delta t$) and short period of analysis,
so the use of price fluctuations or rates of return lead to similar
results. We find that even for low frequency data (large $\Delta t$) 
and for long period of analysis (up to 12 years), the results for the
PDF and the standard deviation remain similar for both 
the price fluctuations and the rates of return [Fig.4].

We have proposed a stochastic process that even in the presence of
correlations among the stochastic variables exhibits a L{\'e}vy
stability for the PDF.  The STL process is characterized by identical
scaling exponents for both the moments and the PDF.  The STL process
provides an unified dynamical picture to describe different
statistical properties, and can be generalized for situations when the
moments and the PDF exhibit different scaling behavior.  The STL
process can be utilized --- as we show in the case for financial data
--- not only for processes with a single scaling regime but also for
physical systems with different regimes of scaling behavior.

\vspace*{-0.5cm}


\vspace*{-1.2cm}
\begin{figure}
\narrowtext
\centerline{
\epsfysize=0.8\columnwidth{\rotate[r]{\epsfbox{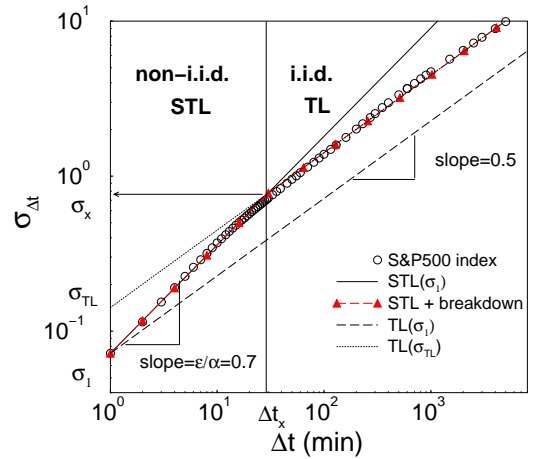}}}}
\vspace*{0.5cm}
\caption{
The $S\&P500$ index shows two regimes of scaling behavior for the
standard deviation $\sigma$.  The correlated (superdiffusive) regime
at small ${\Delta t}$ corresponds to the STL process with slope 
$\epsilon/\alpha=0.7$. To
account for the crossover to uncorrelated (normal diffusion) regime,
we introduce a
breakdown in the scaling for the STL process:
$\lambda_{\Delta t} \equiv \lambda_{\times}=const$ and $A_{\Delta t}
\equiv ({\Delta t}) A_{\times}$ for $\Delta t>(\Delta t)_{\times}$.
The breakdown in the STL is equivalent to a transition to a TL process
at large time scales.  This TL process corresponds to an initial
$\sigma_{TL}$ larger than the empirical $\sigma_1$. This is the reason
for the delay (at time scale $(\Delta t)_s\approx 10^3$) in the
transition from L{\'e}vy to Gaussian behavior observed for ${\cal
P}_{\Delta t}(0)$ [Fig.2]. Note, that the TL process with an initial
standard deviation $\sigma_1$ (as observed in the data) would exhibit
for ${\cal P}_{\Delta t}(0)$ a transition from L{\'e}vy to Gaussian at
shorter time scales.}
\label{fig.1}
\end{figure}

\vspace*{-0.5cm}
\begin{figure}
\narrowtext
\centerline{
\epsfysize=0.8\columnwidth{\rotate[r]{\epsfbox{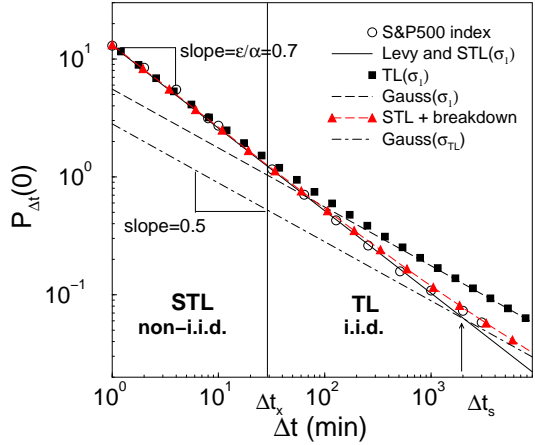}}}}
\vspace*{0.5cm}
\caption{$S\&P500$ data for the 
probability of return to the origin ${\cal P}_{\Delta t}(0)$ 
mimics L{\'e}vy scaling for more than 3 decades in ${\Delta t}$.
The slope and the intercept of the straight line which represents the 
scaling of the L{\'e}vy distribution are determined from the
parameters $\alpha=1.43$ and $A_1=0.0014$  by fitting the initial PDF 
${\cal P}_1(z)$ for the $S\&P500$ index. From the same
fit, we obtain $\lambda_1=0.7$. These initial parameters are used to 
define the STL process. As expected, the STL process follows the 
L{\'e}vy scaling for ${\cal P}_{\Delta t}(0)$ at all time scales.
The TL process (with $\sigma_1=0.07$, identical to the empirical value) 
exhibits a transition at short time scales
to the Gaussian process (with the same value of $\sigma_1$), 
in disagreement with the data. The STL process
with a breakdown at $(\Delta t)_{\times}$, however, is in 
agreement with the data and explains the delayed transition 
(at $(\Delta t)_s\approx 10^3$) to the Gaussian observed in the data. }
\label{fig.2}
\end{figure}

\vspace*{-0.5cm}
\begin{figure}
\narrowtext
\centerline{
\epsfysize=0.8\columnwidth{\rotate[r]{\epsfbox{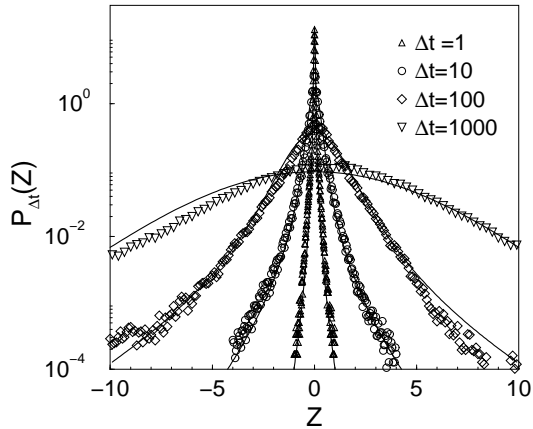}}}}
\vspace*{0.3cm}
\caption{ $S\&P500$ probability distributions ${\cal P}_{\Delta t}(z)$
of price fluctuations $z$ for different time scales $\Delta t$.  With
solid lines, we show the PDF of the STL, with breakdown process for
the same time scales and parameters used in Fig.2. Good agreement
between data and the theoretical PDFs is observed for the central
part. To reproduce better the experimentally observed change in slope
of the far tails we use $\alpha=1.43$, $A_1=0.0028$, and
$\lambda_1=2.6$.  The shape of ${\cal P}_{\Delta t}(z)$ changes as a
function of $\Delta t$ from exponential-like (for small $\Delta t$ ---
STL non-i.i.d. regime) to Gaussian-like profile of the tails (for
large $\Delta t$ --- TL i.i.d. regime).  }
\label{fig.3}
\end{figure}

\vspace*{-0.5cm}
\begin{figure}
\narrowtext
\centerline{
\epsfysize=0.82\columnwidth{\rotate[r]{\epsfbox{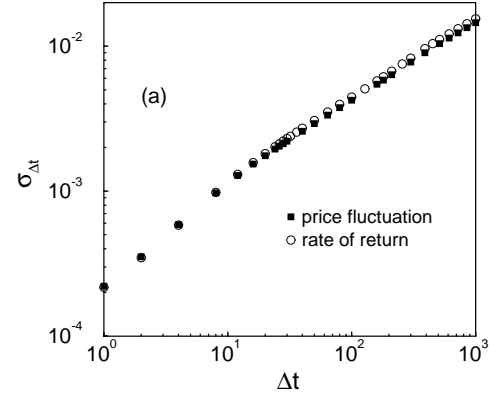}}}}
\vspace*{-0.2cm}
\centerline{
\epsfysize=0.82\columnwidth{\rotate[r]{\epsfbox{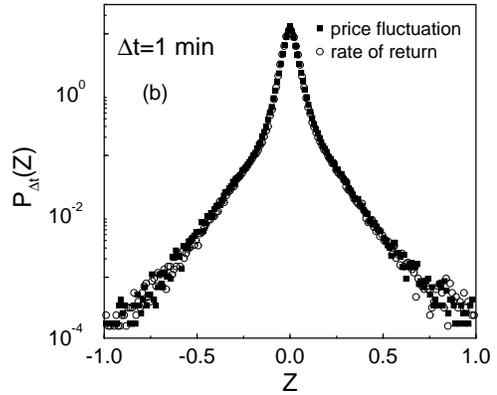}}}}
\vspace*{-0.2cm}
\centerline{
\epsfysize=0.82\columnwidth{\rotate[r]{\epsfbox{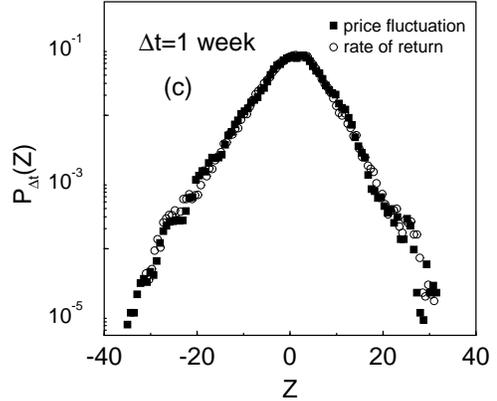}}}}
\vspace*{0.5cm}
\caption{
For the $S\&P500$ index (Jan '84- Dec '95), we show how the standard
deviation of price fluctuations and rates of return completely
overlap, after rescaling by the average price for that period.
Note, that the standard deviation is calculated for a long
period of time (12 years) and for data at either high frequency (1
min) or low frequency (1 trading week). In (b) and (c) we show how the
probability distributions of the same stochastic variables collapse
after rescaling by the same average price. The collapse is
obtained again for both low and high frequency data. We also subdivide
the 12 year period in two equal subintervals, and find the same
collapse. }
\label{fig.4}
\end{figure}

\end{multicols}


\begin{references}
\bibitem{footnote1} 
Different types of smooth truncation can be reproduced with changing
values of $\beta$: (i) for $\beta<1$ we have stretched exponential
tails; (ii) for $\beta=1$ we have an exponential truncation (the
case of \cite{Koponen95}); (iii) $\beta>1$ we have process faster than
exponential; and (iv) for $\beta=2$ we have Gaussian behavior.
\bibitem{footnote3}
Note that the STL process characterized by given $\alpha$ can scale with any
scaling exponent $\epsilon/\alpha$ in contrast to the L\'{e}vy stable
process which scales with the scaling exponent $1/\alpha$. The
parameter $\epsilon$ controls the dynamics of the process ---
probability distributions characterized with the same $\alpha$ can
exhibit different scaling behavior for different values of $\epsilon$.
E.g. for $\epsilon = 1$ and $\lambda = 0$ under the transformations of
Eqs.~(\ref{scale-invar.tr}) the probability density ${\cal P}(z)$
scales as the L{\'e}vy stable process.
\bibitem{footnote4}
We obtain the following analytic expression:
${\cal B}=[\sqrt{2\pi}\sigma_1{\cal L}_{\Delta t=1}(0)]^{2\alpha/(2-\alpha)}$,
where ${\cal L}$ is the L{\'e}vy PDF \cite{Koponen95}.
\end{references}
\end{document}